\documentclass[fleqn,12pt,twoside]{article}
\usepackage[headings]{espcrc1}

\readRCS
$Id: espcrc1.tex,v 1.2 2004/02/24 11:22:11 spepping Exp $
\usepackage{graphicx}
\usepackage[figuresright]{rotating}

\newcommand{\AmS}{{\protect\the\textfont2
  A\kern-.1667em\lower.5ex\hbox{M}\kern-.125emS}}

\title{Theoretical treatments of fusion processes in collisions of
weakly bound nuclei\thanks{This work was supported in part by the CNPq, the
FAPERJ/CNPq(PRONEX) and the FAPESP.}}

\author{
        L.F. Canto\address[UFRJ]{Instituto de F\' \i sica, Universidade 
        Federal do Rio de Janeiro, Cidade Universit\'aria, CT Bloco A, 
        21941-972 Rio de Janeiro, RJ, Brasil},
        R. Donangelo\addressmark\ 
        and
        M.S. Hussein\address{Instituto de F\'\i sica, Universidade 
        de S\~ao Paulo, C.P. 66318, 05389-970, S\~ao Paulo, SP, Brasil}
}
       
\runtitle{Theoretical treatments of fusion processes in collisions of
weakly bound nuclei}
\runauthor{Canto, Donangelo and Hussein}

\begin{document}

% typeset front matter
\maketitle

\begin{abstract}
We review the theoretical methods to evaluate fusion cross sections in
collisions of weakly bound nuclei. We point out that in such collisions the
coupling to the breakup channel leads to the appearance of different fusion
processes. The extension of the coupled-channel method to coupling with the
continuum is the most successful treatment for these collisions. However,
evaluating separate cross section for each fusion process remains a very
hard task.
\end{abstract}

% typeset front matter

\section{INTRODUCTION}

The theoretical description of the fusion of nuclei has so far relied on
a many-degrees-of-freedoms tunneling formalism referred to as the Coupled
Channels Theory. In the fusion of stable nuclei the channels 
considered correspond to bound states in the participating nuclei. When 
weakly bound nuclei fuse, the coupling to continuum channels becomes 
important and one resorts to by and large intuitive means to take these 
into account \cite{CGD06}. If fusion occurs above the Coulomb barrier, 
quantum tunneling is less important and one uses geometrical arguments 
to consider these couplings. In fact this is also done in the description 
of the fusion of fullerene molecules \cite{Cam03}, important in the 
development of strong nano-tubes. At near or sub-barrier
energies, tunneling becomes a dominant feature of the fusion of nuclei and
considering the coupling to the continuum or breakup channels has been a
challenge to theorists. The reason being the need for a full three-body
model for tunneling that describes both the complete fusion as well as the
incomplete fusion of one of the fragments when breakup takes place. In this
three-body model the interactions have to be complex. So far no such
complete theory is practically developed, although serious attempts towards
this goal have been made \cite{MHO04}. Less ambitious endeavours based on 
a discretized Continuum within a basically extended two-body model have 
been extensively used. This is the Continuum Discretized Coupled Channels 
(CDCC) method. Variants of this method have also been pursued especially 
in the fusion of radioactive nuclei. In this contribution we present a 
short account of some of these theoretical attempts to treat the quantum 
tunneling of fragile objects.

\bigskip 

\bigskip 

\section{THE FUSION CROSS SECTION IN COLLISIONS OF STRONGLY BOUND NUCLEI}

Let us first consider the fusion cross section in potential scattering,
where the components of the projectile-target separation vector are the
only degrees of freedom taken into account.
The collision dynamics is described by the Hamiltonian 
\[
H=T+U, 
\]
where $T$ is the kinetic energy operator and $U$ is the optical potential.
In typical situations, one has 
\begin{equation}
U=V+i\left( W^{F}+W^{D}\right) ,  \label{Uopt}
\end{equation}
where $W^{F}$\ and $W^{D}$\ are absorptive potentials, accounting for the
flux lost to fusion and to peripheral processes, respectively. The
scattering state $\left\vert \psi^{(+)}\right\rangle $ is obtained solving
the Schr\"odinger equation with scattering boundary condition and the fusion
cross section is 
\begin{equation}
\sigma_{F}=-\frac{k}{E}~\left\langle \psi^{(+)}\right\vert W^{F} \left\vert
\psi^{(+)}\right\rangle ,  \label{sigF-1ch}
\end{equation}
Above, $E$ is the collision energy and $k$ is the wave number associated to
it.

In multi-channel scattering the problem is more complicated since
the Hamiltonian depends also on intrinsic degrees of freedom. The scattering
wave function is expanded in eigenstates of the intrinsic Hamiltonian, 
$\left\vert \alpha\right\rangle,$ with eigenvalues $\varepsilon_{\alpha}$, in
the form 
\begin{equation}
\left\vert \Psi^{(+)}\right\rangle =\sum_{\alpha}~\left\vert \psi_{\alpha
}^{(+)}\right\rangle ~\left\vert \alpha\right)  \label{psi-expansion}
\end{equation}
and the channel wave functions, $\psi_{\alpha}^{(+)}$, are given by the
coupled-channel equations, 
\begin{equation}
\left[ E_{\alpha}-V_{\alpha}-W_{\alpha}\right] ~\psi_{\alpha}^{(+)}
=\sum_{\beta\left( \neq\alpha\right) }U_{\alpha,\beta}~\psi_{\beta}
^{(+)};\;\;\;\;\;\alpha,\beta=0,1,....~.  \label{CC-eq}
\end{equation}
Above, $E_{\alpha}=E-\varepsilon_{\alpha}$, 
\begin{equation}
V_{\alpha}+W_{\alpha}=\left( \alpha\right\vert U\left\vert
\alpha\right) \;\;\;\;\;\;\mathrm{and}\;\;\;\;\;\;\;U_{\alpha,\beta
}=\left( \alpha\right\vert U\left\vert \beta\right).
\label{potentials}
\end{equation}

If all relevant direct channels are included in the expansion, the imaginary
potential $W_{\alpha}$ represents exclusively fusion absorption. If the
off-diagonal part of the interaction, $U_{\alpha,\beta},$ is real, as
frequently is the case, the fusion cross section can be written 
\begin{equation}
\sigma_{F}=-\frac{k}{E}~\sum_{\alpha}~\left\langle \psi_{\alpha}^{(+)}
\right\vert W^{F}\left\vert \psi_{\alpha}^{(+)}\right\rangle . 
\label{sigfus}
\end{equation}
When the channel coupling interaction is complex, off-diagonal terms should
also be included in the summation above.

It is well known that channel coupling leads to a strong enhancement of the
fusion cross section at sub-barrier energies \cite{DLW83}. That is, the sum
of the contributions from all channels is larger than the fusion cross
section when the coupling is switched off. However, it is important to
investigate the effect of the coupling on the fusion through the elastic
channel alone, $\sigma_{F,0}.$ For this purpose it is convenient to
introduce the polarization potential. Although its calculation may be
difficult, it provides a convenient language for the discussion. This
potential is defined by the condition that the same elastic wave function 
$\psi_{0}^{(+)}$ of the Coupled-channel equations be obtained from a
potential scattering equation (eq. (\ref{CC-eq}) with the RHS set equal to
0), with the replacements 
\begin{eqnarray}
V_{0} & \rightarrow V_{0}+V_{pol}  \label{veff} \\
W_{0} & \rightarrow W_{0}+W_{pol}.  \label{weff}
\end{eqnarray}

The inclusion of $W_{pol}$ always reduces the fusion cross section, since it
leads to absorption of the incident flux before it reaches the region where
fusion takes place. The real part of the potential may lead to additional
fusion suppression, if it is repulsive. However, if it is attractive, it
reduces the barrier height and thus enhances the fusion cross section. In
this case, there is a competition between the real and the imaginary parts
of the polarization potential and the net result depends on the details of
that potential.

\section{FUSION OF WEAKLY BOUND NUCLEI}

Collisions of weakly bound nuclei are much more complicated since the
elastic channel couples strongly with continuum states. In this case, the
channel label becomes continuous and there is an infinite number of coupled
channels, even with truncation at low values of the excitation energy. A
theory including continuum coupling and describing the various fusion
processes is very hard to develop. An important feature of these collisions
is that the breakup mechanism may give rise to two different types of fusion. 
Complete fusion (CF), when the whole masses of the projectile and the target 
are contained in the compound nucleus, and incomplete fusion (ICF), when some 
nucleons move out of the interaction region before the formation of the compound
nucleus. An additional complication is that CF can take place in a single step 
or in a sequential way, as the fragments of the broken projectile are successively 
absorbed by the target. These mechanisms are called Direct (DCF) and sequential 
(SCF) complete fusion, respectively. The sum of CF and ICF is termed total fusion (TF).

In this section, we summarize some of the attempts available in the literature, 
with emphasis on the Continuum Discretized Coupled Channels method (CDCC).

\subsection{Early models}

The earliest calculation of fusion cross sections in collisions of weakly
bound nuclei were carried out with schematic models, based on drastic
approximations. The static effects of the halo could easily be taken into
account through the use of larger diffusivities in the real potential.
However, the inclusion of dynamic effects is a much harder task, since it
involves the coupling with continuum states. 

In the first calculations, coupled channel effects were mocked up by an approximate 
polarization potential. Since the adopted approximations led to a purely imaginary 
potential, changes in the barrier hight were simulated by a
shift in the collision energy. These works found that the CF cross section
was hindered around the barrier energy \cite{HPC92}. Soon after these works, a
calculation adopting the opposite view was performed by Dasso and Vitturi 
\cite{DaV94}. These authors performed coupled channel calculations in  
a simple model in which the continuum was represented by a bound effective channel. 
The CF cross section was then obtained from eq. (\ref{sigfus}), with the exclusion
of the effective channel from the summation. However, in this model 
important properties of the continuum were not properly described. In opposition 
to the previous results, they obtained a strongly enhanced CF cross section.

Since these calculations were based on very schematic models, using
different approximations, it is not surprising that they led to conflicting
conclusions. It is clear that a reliable conclusion about the dynamic
effects of the coupling requires the use of more realistic models.

\subsection{The CDCC method and its applications}

The CDCC method consists of replacing the continuous part of the projectile's 
spectrum by a finite number of discrete states. This can be done in different 
ways \cite{CDCC}. The most frequent approach is to divide the continuum in bins, 
distributed up to some cut-off energy $E_{max}$, and to build wave 
packets within each bin. The wave packets are superpositions of energy eigenstates 
$f_\alpha(k,r)$, where $k$ is the wave number associated with the relative motion 
of the projectile's fragments and $\alpha$ stands for the remaining projectile's 
quantum numbers. One common option is to use $N$ bins of equal size, $\Delta$, 
in momentum space, distributed between 0 and $k_{max}=N\Delta$. 
In this case, the $i^{th}$ discrete state with quantum numbers $\alpha$ is
\begin{equation}
u_{\alpha,i}(r)=\int_{k_i-\Delta/2}^{k_i+\Delta/2}\ g_\alpha (k-k_i)\
f_\alpha(k,r) \ dk.  \label{wpacket}
\end{equation}
The particular shapes of the wave packets do not significantly affect the
results. However, they should be chosen such that the states $u_{\alpha,i}(r)$ 
form an orthonormal set. In this way the problem is reduced to a set of 
coupled-channel equations involving the elastic, the excited bound channels and 
$N$ discretized channels for each $\alpha$. 

An example of the application of this method to the $^8$B breakup process may be 
found in ref.~\cite{TNT}. The comparison of the full CDCC calculations with other 
simplified calculations shows the importance of higher-order couplings,
and also of the couplings between continuum states.

\begin{figure}[htb]
\begin{minipage}[t]{80mm}
\includegraphics*[width=8cm]{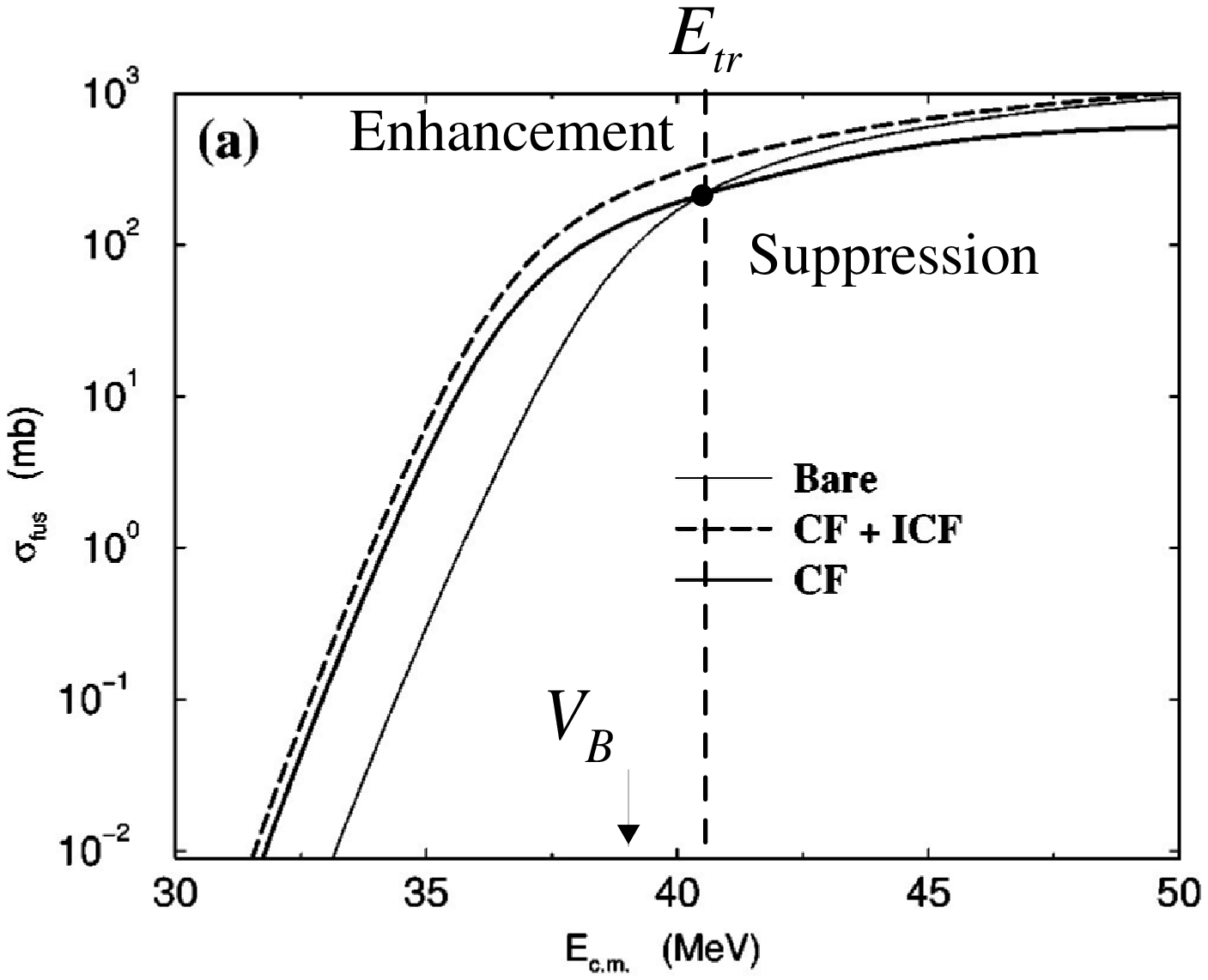}
\caption{CDCC calculation of the fusion cross sections without
continuum-continuum couplings (figure from ref. \cite{HVDL}) }
\label{fig1}
\end{minipage}
\hspace{\fill}
\begin{minipage}[t]{75mm}
\includegraphics*[width=6cm]{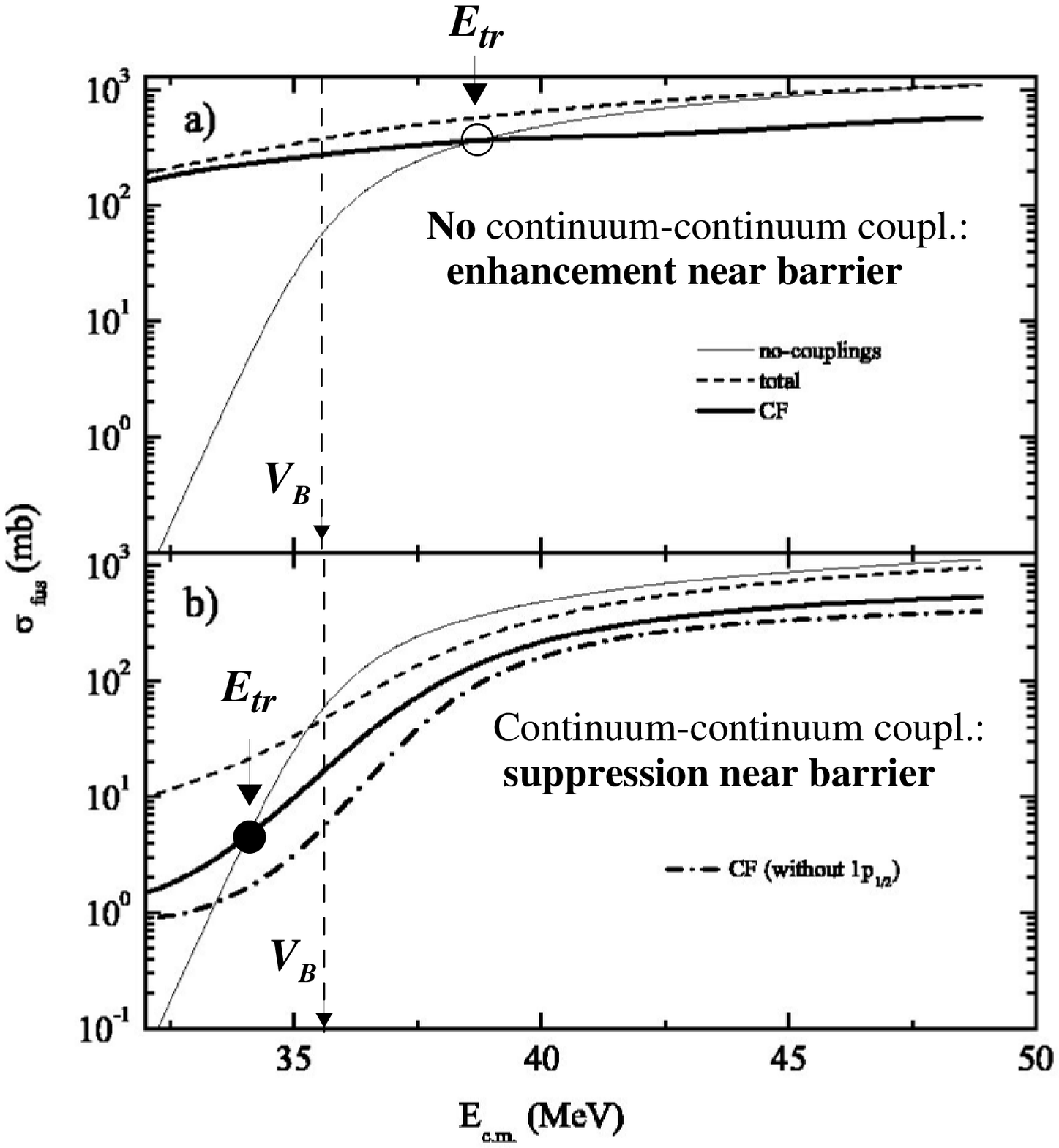}
\caption{CDCC calculation of the fusion cross sections with
and without continuum-continuum couplings (figure from ref. \cite{DiT02}).}
\label{fig2}
\end{minipage}
\end{figure}

This last ingredient, continuum-continuum couplings, had not been included
in the calculations of the fusion cross section for the $^{11}$Be + $^{208}$
Pb collision presented in ref.~\cite{HVDL}. Their results are shown in 
figure \ref{fig1}, in comparison with predictions of a bare potential (channel-coupling 
switched off). Within those conditions, the CF cross section is strongly enhanced
at low energies and slightly hindered at high energies. The transition between
these behaviors occurs at an energy $E_{tr}>V_B$. Thus, this study leads
to enhancement of the CF cross section in the barrier region. 

The effect of the continuum-continuum couplings on this reaction was
investigated by Diaz-Torres and Thompson \cite{DiT02}. Their results are 
shown in figure \ref{fig2}. Continuum-continuum couplings were not included
in the CDCC calculations presented in (a) and were included in the ones shown in 
(b). In (a), the results are similar to those of ref.~\cite{HVDL} while
in (b) they are rather different. Both the CF and TF cross sections are much 
smaller than those obtained without continuum-continuum couplings. In this 
case, the CF cross section is suppressed in the barrier region, with $E_{tr}$ 
falling below the Coulomb barrier. It should be remarked that neither of 
these works take into account the sequential capture of all fragments, and thus 
they provide only a lower limit for the CF cross section. 

Intuitively, the reduction of the fusion cross section arising from couplings
among many channels in the continuum sounds reasonable. It is, however, of interest 
to prove this formally. One can  formulate the problem using Feshbach theory and 
find a way of eliminating the continuum-continuum coupling terms. What one finds is a 
set of coupled channels equations of the type:
\begin{eqnarray}
\left(E_0-H_0 \right)\Psi_0 &= \sum_c\ V_{0c}\Psi_c \\
\left(E_c-H_c \right)\Psi_c &= F_{c0}\Psi_0
\end{eqnarray}
where $V$ and $F$ are different operators. Now we know from compound nucleus theory
that if this happens (the c-channels are closed here) then time reversal is
violated according to the definition of Wigner of time reversal invariance 
\cite{FHK95}. If time reversal is violated one has irreversibility. However, what 
we are interested here is the so-called statistical mechanics irreversibility. Of 
course, in order for this to happen, one has to resort to coarse graining the 
continuum-continuum couplings (energy-averaging) as it has been done extensively in 
the past in the treatment of Deep Inelastic Collisions \cite{KAW79}. This may 
indeed be happening in the actual calculations using the CDCC method.

New calculations employing the CDCC method to determine the effect of the
breakup channel on the total fusion cross section in reactions induced by 
$^{6,7}$Li projectiles were performed more recently \cite{DiTB}. The main
difference between these calculations and those of ref. \cite{DiT02} is in
the absorptive interaction. Ref.~\cite{DiTB} uses a projectile-target imaginary
potential while ref.~\cite{DiTB} considers absorption of each projectile's 
fragment separately. In this way, only the TF cross section is evaluated. 
It was found through those calculations \cite{DiTB} that the presence of the
breakup channel enhances the total (complete + incomplete) fusion cross
section in the region around the Coulomb barrier, and it has very little
effect at higher energies. It was further observed that the effect depended
on the target size, as expected, being more marked in the case of a $^{209}$%
Bi than for $^{59}$Co. The differences in total fusion cross section between
the two Lithium projectiles may be attributed to the coupling to the breakup
channel in the case of $^{59}$Co, but not for the $^{209}$Bi target.

One should also mention the combined CDCC-Dynamic Polarization Potential
(DPP) approach introduced by Keeley, Rusek and collaborators \cite{RKKR,RKK03}.
The procedure consists in first performing CDCC calculations to describe the
elastic scattering data for the system under consideration, and from the calculations
extract the corresponding polarization potentials. The fusion cross sections
are then obtained using a simple barrier penetration calculation, where the
barrier is given by the addition of the real parts of the optical potential 
and the polarization potential.

\begin{figure}[ptb]
\begin{center}
\includegraphics*[width=10cm]{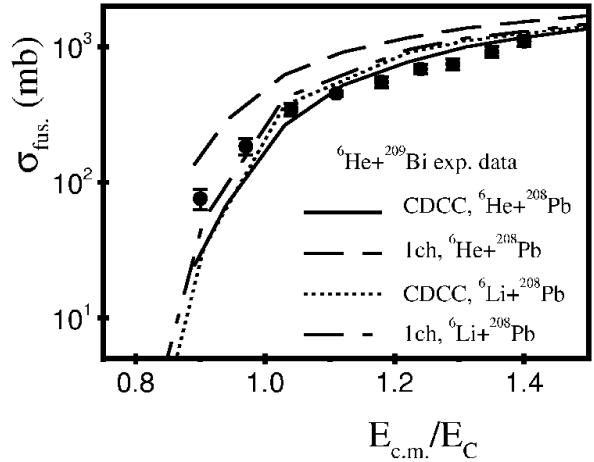}
\end{center}
\caption{CDCC and single-channel calculations of fusion cross sections for
the $^{6}$He + ${208}$Pb and $^{6}$Li + $^{208}$Pb systems. Data for $^{6}$%
He + ${209}$Bi are also shown (figure from ref. \protect\cite{RKK03}).}
\label{fig3}
\end{figure}

This CDCC-DPP method was employed to extract the effect of the $^6$He dipole
polarizability on its elastic scattering on $^{208}$Pb, and its
correspondent effect on the fusion cross section for this reaction. A
comparison of the theoretical fusion cross section of the projectiles $^6$He
and $^6$Li incident on $^{208}$Pb is shown in fig.\ref{fig3}. As there is no
available data for these systems, the calculations are compared with data for
the $^6$He on $^{209}$Bi fusion cross section. The figure shows results for these
projectiles with and without coupling to the breakup channel. When this coupling
is left out (only static effects), the cross section for $^6$He is much larger
than that for $^6$Li, owing the $^6$He neutron halo. When breakup couplings are 
taken into account, repulsive potentials reduce the cross sections. Since this 
effect is stronger for $^6$He, the static effect of the halo is cancelled and the 
fusion cross section for the two projectiles are similar. The CDCC cross section 
for $^6$He projectiles is consistent with the data.

The breakup of the $^6$He and $^6$Li projectiles proceeds according to
different mechanisms. In the case of $^{6}$He, the $E$1 transitions to the
continuum dominate. As the $^4$He and $^2$H fragments of $^6$Li have the
same charge/mass ratios, these transitions do not play a role in the case of 
$^6$Li, which breakup is thus produced by quadrupole couplings, mainly driven
by nuclear forces.

\subsection{Other approaches}

Besides the quantal coupled-channels procedures mentioned above, other
alternative methods have been employed to describe the fusion and breakup
processes in reactions induced by weakly bound projectiles. Recently, Hagino
and collaborators have developed a classical trajectory method, in which the
classical trajectories for the target and the two projectile fragments are
evaluated for a Monte Carlo sampling of initial conditions \cite{HDH}. This
method automatically separates the complete and incomplete fusion cross
sections, so it does not suffer from the deficiency of the CDCC calculations
presented in \cite{DiT02} which count the sequential fusion of all fragments
as an incomplete, rather than a complete fusion process. In fact, according
to these classical calculations, the cross section for the capture of all
fragments represents an important contribution to the complete fusion cross
section. The results of these calculations are in reasonable agreement with
the experimental data at energies above the Coulomb barrier. However, as
they do not include tunneling effects, they cannot be employed at the
underbarrier energies where much of the interest in fusion induced by weakly
bound projectiles lies.

A time-dependent description of the collision process was developed to study 
fusion reactions induced by halo nuclei \cite{YIKUN}. In this approach the 
system is described as a three-body system, composed of the target, the 
projectile core and the halo nucleon. This three-body reaction problem is
described in terms of the time evolution of a wave packet. The initial wave
packet is taken to be the product of an incoming wave for the relative motion 
between the centers of mass of the projectile and target nuclei, multiplied by 
a Gaussian wave packet representing the initial configuration of the projectile 
fragments. The evolution of this wave packet is treated through the standard
time-dependent Schr\"odinger equation approach, which although requiring long
calculation times, is simple to program. The fusion probability is calculated 
through an energy projection procedure. This approach has been applied to several
systems \cite{IYNU}. In the case of neutron halos, it is found that the effect 
of the halo hinders the fusion cross section.

Since semiclassical calculations of continuum discretization coupled-channels
have been successfully applied to the description of the breakup of weakly bound 
projectiles \cite{MCDL}, they should be extendable to the treatment
of fusion reactions. The breakup amplitudes calculated with the semiclassical
calculations would determine the initial conditions for the study of the 
fragments. Such a procedure would have the advantage of automatically separating
the motion of all fragments, thus separating the sequential fusion contribution
to complete fusion from the incomplete fusion process, which was one of the
difficulties facing the quantum mechanical CDCC calculations. A preliminay calculation
of the CF cross section in a schematic two-channel model has recently been performed
\cite{CDM06} and the results were compared with results of the coupled channel method.
The agreement was good.

\section{SUMMARY}

In this paper we have presented a brief review of the theoretical methods to describe
fusion reactions with weakly bound nuclei. We pointed out that these reactions
involve the coupling to continuum states, which is very hard to describe. In
this way, simple models using drastic approximations can hardly provide a reasonable
description of the fusion processes. The most successfull calculations performed
so far are based on the CDCC method. The major lesson we have learnt from them
is that continuum-continuum couplings play an essential role in the fusion processes,
reducing substantially the CF cross section. This result arises from the 
statistical irreversibility of transitions to the continuum. We have pointed
out that the presently available CDCC calculations have the shortcoming of not 
providing separate descriptions of the various fusion processes following breakup.
This is a very hard task, which might be handled in a better way in classical and
semiclassical models.


\begin{thebibliography}{99}
\bibitem{CGD06} L.F. Canto, P.R.S. Gomes, R. Donangelo and M.S. Hussein,
                Phys. Rep. 424 (2006) 1.

\bibitem{Cam03} E.E. Campbell, Fullerene Collision Reactions, 
                Kluwer Academic Publishers, Dordrecht, the 
                Netherlands (2003).

\bibitem{MHO04} T. Matsumoto, E. Hiyama, K. Ogata, Y. Iseri, M. Kamimura, 
                S. Chiba and M. Yahiro, Phys. Rev. C70 (2004) 061601.

\bibitem{DLW83} C.H. Dasso, S. Landowne and Aa. Winther, Nucl. Phys.  
                A405 (1983) 381; A407 (1983)221; W. Reisdorf \textit{et al.}, 
                Nucl. Phys. A614 (1997) 112.

\bibitem{CDCC}  see e.g. N. Austern \textit{et al.}, Phys. Rep. 154 (1987)
                125  and references therein; F. P\'eres-Bernal 
                \textit{et al.},  Phys. Rev. A63 (2001) 052111.

\bibitem{HPC92} M.S. Hussein, M.P. Pato, L.F. Canto and R. Donangelo, 
                Phys. Rev. C46 (1992) 377; N. Takigawa, M. Kuratani 
                and H. Sagawa, Phys. Rev. C47 (1993) R2470.

\bibitem{DaV94} C.H. Dasso and A. Vitturi, Phys. Rev. C50 (1994) R12.

\bibitem{TNT}   J.A. Tostevin, F.M. Nunes, and I.J. Thompson, Phys. Rev. 
                C 63 (2001) 024617.

\bibitem{HVDL}  K. Hagino, A. Vitturi, C.H. Dasso and S. Lenzi, Phys. 
                Rev. C61 (2000) 037602.

\bibitem{DiT02} A. Diaz-Torres and I.J. Thompson, Phys. Rev. 
                C65 (2002) 024606.

\bibitem{FHK95} H. Feshbach, M.S. Hussein and A.K. Kerman, Z. Phys. A351 
                (1995) 133.

\bibitem{KAW79} C.M. Ko, D. Agassi and H.A. Weidenm\"uller, Ann. Phys. 117 (1979) 237.    
\bibitem{DiTB}  A. Diaz-Torres, I.J. Thompson and C. Beck, Phys. Rev. 
                C 68 (2003) 044607.

\bibitem{RKKR}  N. Keeley and K. Rusek, Phys. Lett. B 427 (1998) 1;  
                N. Keeley, K.W. Kemper and K. Rusek, Phys. Rev. C65 (2001) 
                014601;  ibid. Phys. Rev. C66 (2003) 044605(R);  

\bibitem{RKK03} K. Rusek, N. Keeley, K.W. Kemper and R. Raabe, Phys. Rev. 
                C67 (2003) 041604(R).

\bibitem{HDH}   K. Hagino, M. Dasgupta and D.J. Hinde, Nucl. Phys. 
                A 738 (2004) 475.

\bibitem{YIKUN} K. Yabana, M. Ito, M. Kobayashi, M. Ueda and T. Nakatsukasa, 
                Nucl. Phys. A 738 (2004) 303. 

\bibitem{IYNU}  M. Ito, K. Yabana, T. Nakatsukasa and M. Ueda, 
                arXiv:nucl-th/0506073. 

\bibitem{MCDL}  H.D. Marta, L.F. Canto, R. Donangelo and P. Lotti, 
                Phys. Rev. C 66 (2002) 83.

\bibitem{CDM06} L.F.Canto, H.D. Marta and R. Donangelo, Phys. Rev. C 73 (2006) 034608.

\end{thebibliography}
\end{document}